\documentclass[
 reprint,
 amsmath,amssymb,
 aps,
 prf
]{revtex4-1}

\usepackage{graphicx}
\usepackage{dcolumn}

\usepackage{amssymb}
\usepackage{amsmath}
\usepackage{mathtools}

\begin{document}


\title{Contact line depinning from sharp edges.}


\author{J. Gra\~na Otero}
\email{jose.grana@uky.edu}
\affiliation{279 RGAN. Dept. of Mechanical Engineering. University of Kentucky, KY, USA}
\author{I. E. Parra Fabi\'an}
\affiliation{Center for Computational Simulation (CCS). Universidad Polit\'ecnica de Madrid. Espa\~na}


\date{\today}

\begin{abstract}

With aim of finding mathematical criteria for contact line depinning from sharp corners, we have studied the equilibrium and stability of a semi-infinite planar liquid layer pinned at the vertex of a wedge. Equilibrium is compatible with a fan of apparent contact angles $\theta_0$ bracketed by the equilibrium contact angles of both flanks of the wedge, so the contact line could remain pinned if $\theta_0$ is within this fan. However, the analysis of the stability of these solutions, studied exploiting the variational structure of the problem through turning-point arguments, shows that the equilibrium becomes unstable at critical depinning advancing $\theta_0^a$ and receding $\theta_0^r$ contact angles, which are found as subcritical saddle-node bifurcations. Equilibrium is thus possible (stable) within the interval $\theta_0^a < \theta_0 <\theta_0^a$ but the contact line depins from the vertex beyond these critical angles if they occur within the equilibrium fan.

\end{abstract}

\pacs{47.55.np, 47.61.-k}
\keywords{Contact line, Pinning, Superhydrophobicity, Bond number, Eigenvalue problems}

\maketitle


\section{Introduction}

Static liquid free interfaces meet smooth solid surfaces along the contact line, also called triple line, with a well-defined (uniquely determined) angle, the so-called contact angle, which is a thermodynamic property at equilibrium\cite{deGennesReview, Adamson, deGennesBook}. If instead the triple line glides over the surface, the contact angle is no more a thermodynamic property, although it is still characterized by well-defined advancing and receding values, typically functions of the instantaneous apparent relative velocity\cite{BlakeReview, Eggers, ARMoving, Hysteresis}.

However, the contact angle is not uniquely determined by a local equilibrium analysis on surface singularities, i.e. points or lines where the vector normal to the surface is not well-defined, such as sharp corners. Instead, in these cases a fan of angles is compatible with the pinned triple line, so it can pivot around the singularity while still pinned, even when the liquid and its free interface around the triple point are in motion\cite{MaxPlank, EnergyV, ScienceBouncingDrops, DropBouncing}. For instance, in the case of the planar configuration of Fig.~\ref{OverhangingDrop}, with an interface pinned at the vertex of a wedge so the triple line is just a point, it was first shown by Gibbs\cite{Gibbs}, and later by others both experimentally and theoretically \cite{Dyson, GibbsCondition, HuhMason, MasonZero}, that, at equilibrium, the contact angle can span the fan defined by the equilibrium contact angles of the two flanks forming the wedge.

Triple line configurations with more complex geometry than that considered by Gibbs are considerably more difficult to analyze. This is the case for instance of  the interaction of contact lines with superhydrophobic substrates\cite{QuereReview, Thermite}. This type of surfaces is typically characterized by complex micro-roughness composed of discrete structures such as posts with a variety of cross section shapes, and with characteristic scales much smaller than the capillary length\cite{ DualScale, BHUSHAN2011, Extrand, Lotus, Hierarchical, HierarchicalII, Selfhealing, Plants, Ice, SelfSimilar, HierarchicalReview, HierarchicalIII}. In these cases, the contact line has typically two different length scales. Locally close to the solid substrate, the interface is deformed and develops wrinkles with the same characteristic length as the micro topology. However, at the macroscopic scale, for instance the diameter of the drop or the thickness of the liquid layers resting on the substrate, the interface can be characterized by well-defined apparent contact angles\cite{EnergyII, Review}.

Recently, studies solving the details of the geometry of the interface around the contact line in complex configurations have been published\cite{ComplexShapes, ComplexShapesI, ComplexShapesII, ComplexShapesIII, ComplexShapesIV,PinningWang}. However, the more common approach is to use homogenized mesoscopic analyses\cite{EnergyV, Review, Energy, EnergyIII, EnergyII}, often based on energy arguments, which somehow circumvent the difficulties associated with the detailed description of the interface. Important contributions from this point of view are the studies due to Pomeau and Vannimenius\cite{Pomeau} and to Joanny and de Gennes\cite{GennesDefects}. Both analyzed the limit of weak heterogeneities on the substrate, the former in static conditions, the latter in the case of a moving triple line pinning on an isolated defect.

Two other seminal contributions using homogenized viewpoints are the phenomenological and now classical models due to Cassie-Baxter\cite{CassieBaxter} and to Wenzel\cite{Wenzel}, originally proposed to address the specific case of superhydrophobic substrates. Wenzel's model is appropriate for the so-called wetting conditions, when the liquid wets the entire exposed solid surface; in this case the apparent contact angle at the triple line is given in terms of the ratio of the wetted to the projected areas of the solid surface. Cassie-Baxter's model applies instead to the so-called non-wetting conditions, when the liquid entraps underneath its free surface gas pockets which prevent its direct contact with the solid; in this case, the apparent contact angle is given in terms of ratio of the wetted to the exposed areas of the solid surface. 

These mesoscopic homogenized models provide guidance understanding experimental work. However, precise quantitative criteria predicting depinning still remains vague despite their critical role in phenomena involving liquid free interfaces in contact with solids. For instance, the transition between wetting and non-wetting states on superhydrophobic surfaces is well-known to be mediated by depinning transitions, both in static\cite{SuperHBreakdown}, and in the dynamics conditions\cite{MaxPlank}. Or in boiling liquids, vapor bubbles grow on microscopic wall cavities up to final bubble sizes at detachment which depend critically on whether the triple line of the bubble remains pinned at the rim or it glides and spreads over the wall around the cavity\cite{Attinger, BoilingContactAngle, BoilingContactAngleII, BubblePinningII, BubblePinningIII, Reentrant}.

This paper is devoted to the mathematical analysis of depinning from sharp edges. We show how depinning criteria can be precisely defined, illustrating the methodology in a simple configuration so the mathematical treatment remains affordable. The essential idea is to formulate the problem for the interface shape as an eigenvalue problem, with the Bond number $Bo$ as the eigenvalue in the specific case studied. Depinning can be found then as a subcritical saddle-node bifurcation for a critical Bond number. This basic idea is still generally applicable in more complex configurations despite the considerable mathematical difficulties.

\section{Advancing triple lines.} 

\begin{figure}[h!]
	\includegraphics[width=\linewidth]{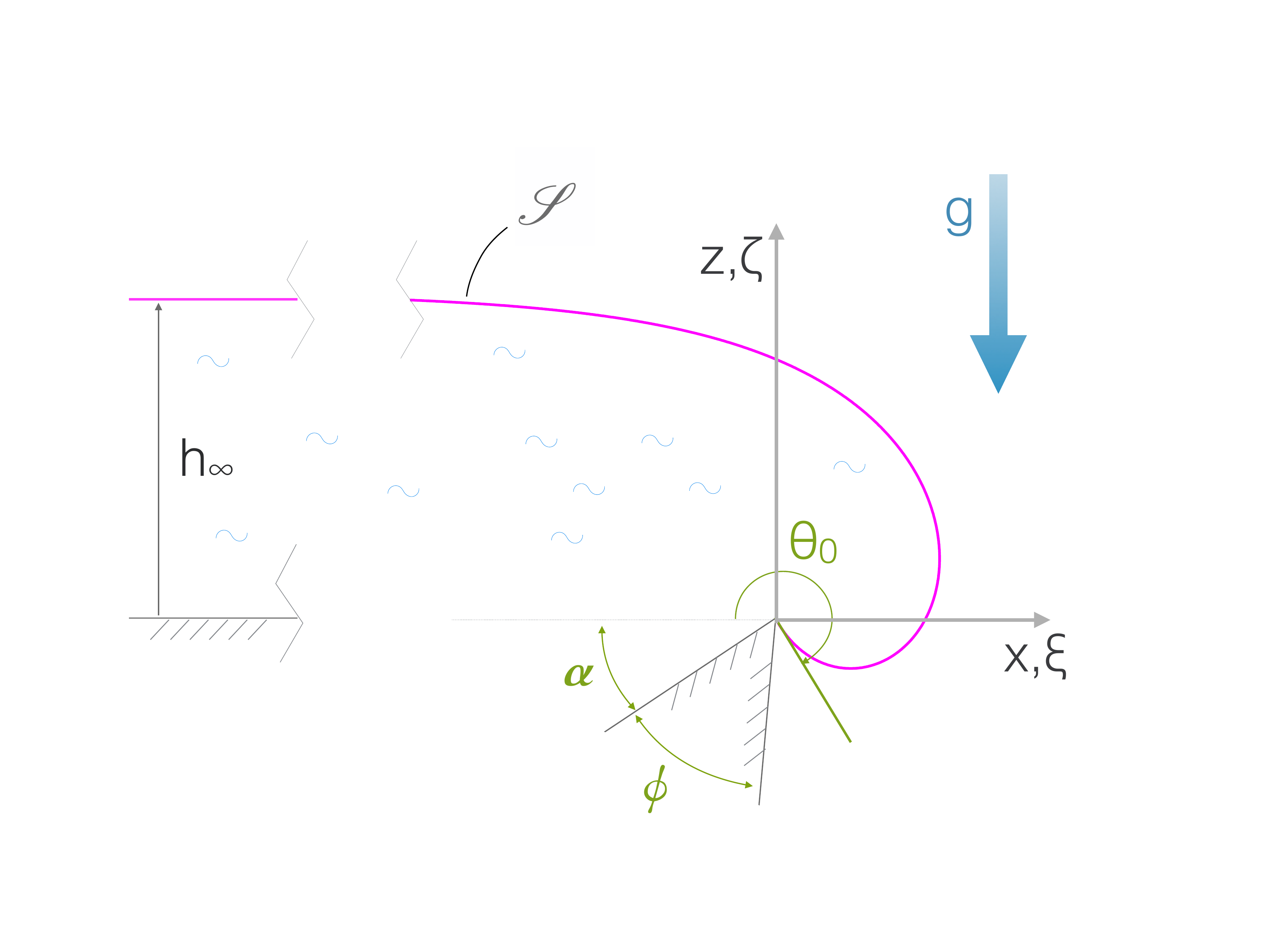}
	\caption{Geometry of the liquid layer interface (pink line) pinned at the vertex of a wedge of inner angle $\phi$. The plotted interface is the equilibrium shape corresponding to an angle $\theta_0 = 240^\circ$, or equivalently to a Bond number $Bo = \rho g h_\infty^2/\sigma = 3$, i.e. a thickness upstream $h_\infty =  \sqrt{3}\ell_c$ with $\ell_c^2 =  \sigma/(\rho g)$ the capillary length. As shown below, this is an equilibrium but unstable configuration.}
	\label{OverhangingDrop}
\end{figure}

Figure \ref{OverhangingDrop} shows a sketch of the 2D configuration we study here, namely a semi-infinite horizontal liquid layer in static equilibrium under the action of gravity and surface tension forces and pinned at the vertex of a wedge. The material of the front and back flanks of the wedge can be different, and the corresponding equilibrium contact angles are $\theta_{ef}$ and $\theta_{eb}$ respectively. In order to ensure equilibrium, the solid substrate must become horizontal (perpendicular to gravity) far upstream from the wedge, so the liquid layer also levels off to a constant thickness $h_\infty$, which ultimately sets the pressure distribution inside the liquid layer. The actual details of the substrate between the wedge and this horizontal zone are actually irrelevant in this static problem. 

Initially the interface meets the smooth substrate upstream from the wedge's vertex with the corresponding equilibrium contact angle $\theta_0 = \theta_{ef}$. In these conditions, as shown below, the balance between gravity and surface tension uniquely determines\cite{deGennesBook} the liquid layer thickness $h_\infty$ in terms of $\theta_{ef}$ as:
\begin{gather}
	\label{EqHeight}
	h^2_\infty = 2(\sigma/\rho g) (1 - \cos\theta_{ef}).
\end{gather}
Angles are measured through the liquid, relative to the local horizontal, positive clockwise as shown in Figure \ref{OverhangingDrop}.

\begin{figure}[h!]
	\includegraphics[width=\linewidth]{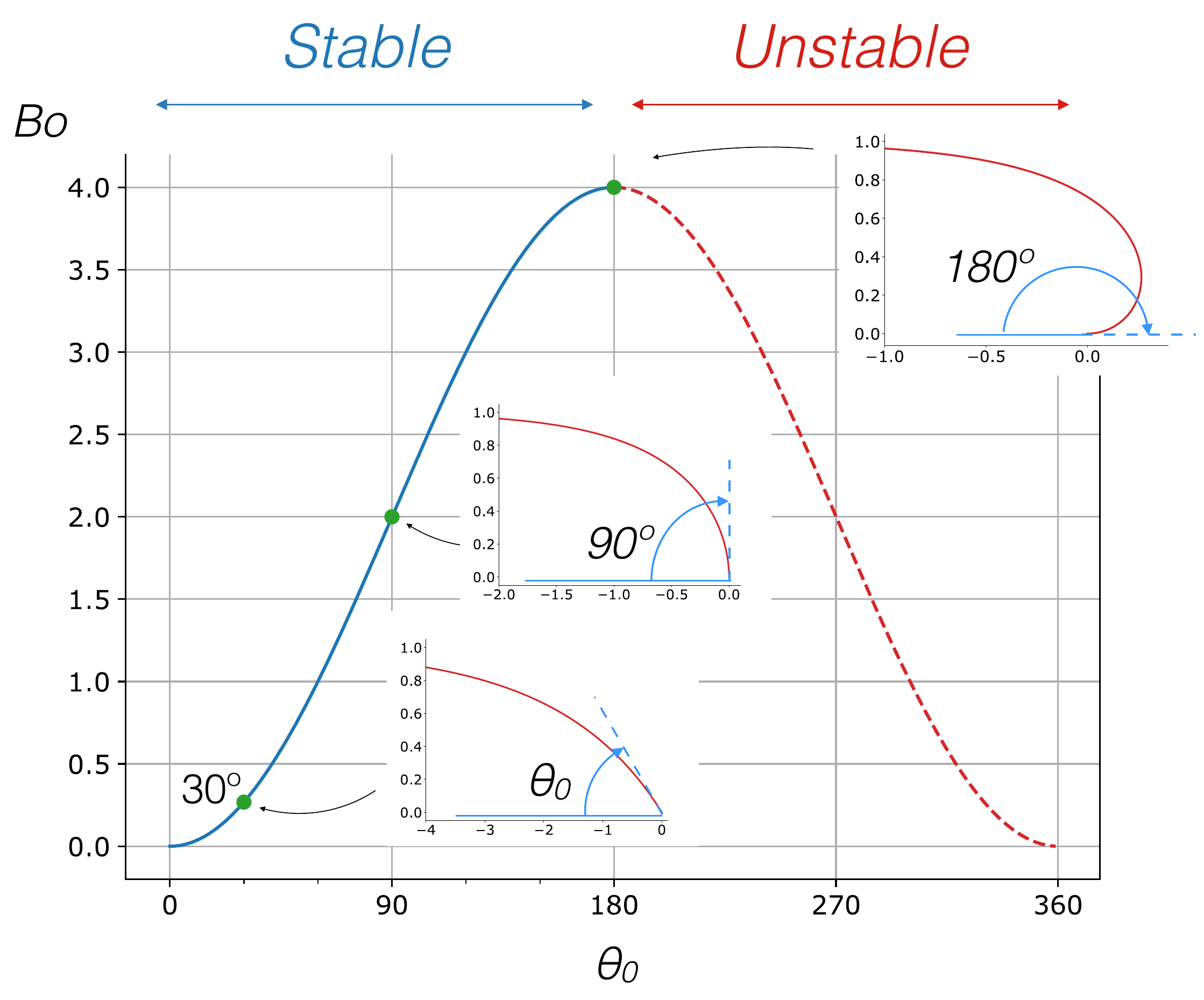}
	\caption{(Color online) The continuous blue and dashed red lines represent the equilibrium (thickness) Bond number $Bo = \rho gh_\infty^2/\sigma$ as a function of the apparent angle $\theta_0$ at the edge, as given by the expression \eqref{BoTheta}. The blue line represents the stable equilibrium solutions branch, whereas the red dashed line are unreachable equilibrium solutions because depinning occurs before. The inserts show the shape of the interfaces for some representative values of $\theta_0$ on the stable branch. }
	\label{toBo}
\end{figure}

The liquid layer is then forced quasi-statically to slowly glide --  for instance by continuously supplying liquid far upstream from the wedge -- over the front flank of the wedge until the triple line touches the vertex. Pushing the liquid layer further against the vertex forces the contact angle $\theta_0$ to increase beyond $\theta_{ef}$. With the triple line pinned, the angle $\theta_0$ is now a free parameter, which is no longer determined by thermodynamic equilibrium. Instead, the liquid layer thickness $h_\infty$ and the angle $\theta_0$ are now arbitrary, only linked by the balance between gravity and surface tension forces which leads to the same relationship as in \eqref{EqHeight} but with $\theta_0$ replacing $\theta_{ef}$. In terms of the capillary length $\ell^2_c = \sigma/(\rho g)$ and of the Bond number $Bo = (h_\infty/\ell_c)^2$ this expression can be alternatively written as:
\begin{gather}
	\label{BoTheta}
	Bo = \left(\frac{h_\infty}{\ell_c}\right)^2 = 2(1 - \cos\theta_0)
\end{gather}
which is represented in Figure \ref{toBo}.

Thus, forcing the liquid layer against the wedge's vertex, as trying to overcome it, requires at equilibrium to increase the thickness $h_\infty$ of the liquid layer, and accordingly the angle $\theta_0$, as Eq. \eqref{BoTheta} shows. The triple line can be expected to still remain pinned at the vertex until, in principle, the equilibrium contact angle with the back flank is reached. Once this condition is reached, the triple line would continue gliding over this flank. The apparent contact angle $\theta_0$ spans therefore, with the triple line pinned, a fan of angles bracketed by the lower $\theta_0^{min} = \theta_{ef} - \alpha$ and the upper $\theta_0^{max} = \pi - \alpha - \phi + \theta_{eb}$ limits, as Gibbs first and others later showed\cite{Gibbs,Dyson,GibbsCondition}. 

However, as we show below, the interface becomes unstable at a critical angle $\theta_0^{a}$ (the superscript $a$ stands for advancing front). In this case, the triple line will depin if $\theta_0^{a}$ is contained within the Gibbs' fan. More specifically, for the configuration considered here the interface destabilizes at $\theta_0^{a} = \pi$ so if $\theta_0^{min} < \pi < \theta_0^{max}$, then the interface depins before starting to glide along the back surface. This depinning condition is suggested by Figure \ref{toBo}, which shows that the height $h_\infty$, or equivalently the Bond number $Bo$, as functions of the angle $\theta_0$, exhibit a maximum defined by:

\begin{gather}
	Bo^{a} = 4 \qquad \text{for} \qquad \theta^a_0 = \pi,
\end{gather}
or equivalently $h_\infty^{a} = 2\ell_c$.

Thus, forcing more liquid into the liquid layer, trying to increase its thickness and the contact angle in order to overcome the wedge's edge, ultimately fails at the critical angle $\theta^a_0 = \pi$ because equilibrium is not any more possible for thicknesses greater than $h^{a}_\infty$. The triple line depins therefore. This argument can be made rigorous by studying the stability of the equilibrium solutions to show that $\theta^a_0 = \pi$ actually represents a bifurcation point where the equilibrium solutions for $\theta^a_0 > \pi$ become unstable as shown in Section \ref{AppStability}.

\subsection{Formulation.}\label{Formulation}

The analysis has been straightforward so far due to the simple configuration considered. However, a more general and systematic approach will be used below, in order to show how to analyze more complex geometries. At equilibrium, the pressure distribution in the liquid is given by the hydrostatic balance $p(z) = p_\infty + \rho g (h_\infty - z)$, so the shape of the interface $\mathcal{S}$ is determined by the condition that $p(z)$, evaluated at $\mathcal{S}$, is $p(\mathcal{S}) = p_\infty + \sigma \mathcal{K}$, where $\sigma$ is the surface tension of the liquid in contact with the ambient gas, and $\mathcal{K}$ is twice the mean curvature of the interface.

The geometry of the liquid interface will be represented parametrically in the form $(x_\mathcal{S}(t), z_\mathcal{S}(t))$, where $x_\mathcal{S}$, $z_\mathcal{S}$ are, respectively, the horizontal and vertical coordinates of the interface in the system of reference of Figure \ref{OverhangingDrop}. The parameter $t$ is the arc length, so $x_\mathcal{S}'^2(t) + z_\mathcal{S}'^2(t) = 1$, with the primes representing the $t$-derivative. Lengths are made dimensionless with the far-upstream liquid thickness $h_\infty$, leading to the dimensionless variables $\xi = x_\mathcal{S}/h_\infty$, $\zeta = z_\mathcal{S}/h_\infty$ and $s = t/h_\infty$; the s-derivative is represented below with a dot, $\dot{\phi} = d\phi/ds$. With this parametrization, the dimensionless interface curvature can be written as\cite{doCarmo,Eisenhart} $\kappa_\mathcal{S} = h_\infty\mathcal{K} = \dot{\xi}_\mathcal{S} \ddot{\zeta}_\mathcal{S} - \dot{\zeta}_\mathcal{S} \ddot{\xi}_\mathcal{S}$, so the shape of the interface is determined by the fourth order differential system (dropping for simplicity the subscript $\mathcal{S}$):
\begin{subequations}
	\label{ProblemParametricF}
	\begin{gather}
		\label{EqsDiff1}
		\dot{\xi} \ddot{\zeta} - \dot{\zeta} \ddot{\xi} = Bo(1 - \zeta)\\
		\label{EqsDiff2}
		\dot{\xi}^2 + \dot{\zeta}^2 = 1
	\end{gather}

	The appropriate boundary conditions are those enforcing pinning at the edge of the substrate:
	\begin{gather}
		\xi(s = 0) = \zeta(s = 0) = 0,
	\end{gather}
	and leveling off to $h_\infty$ far upstream from the edge:
	\begin{gather}
		\xi \rightarrow -\infty, \quad \zeta \rightarrow 1 \qquad \text{as}  \qquad s \rightarrow \infty.
	\end{gather}
\end{subequations}

This problem has a continuum of solutions parametrized by the Bond number, which enters due to the presence of two characteristic lengths, namely the liquid layer thickness $h_\infty$, and the capillary length $\ell_c$. In particular, the apparent contact angle $\theta_0$ of the interface at the edge can then be determined from either $\cos\theta_0 = -\dot{\xi}(0)$ or $\sin\theta_0 = \dot{\zeta}(0)$, which give $\theta_0$ as a function of $Bo$ as shown in Figure \ref{toBo}. This Figure reveals that $\theta_0$ is a multi-valued function of the Bond number $Bo$. On the contrary, $Bo$ is univocally determined as a function of $\theta_0$. Thus, the problem of finding the shape of the interface is more conveniently formulated as an eigenvalue problem, i.e. that of finding the Bond number $Bo$ that corresponds to a given angle $\theta_0$ at the edge. In this case, the boundary conditions must be augmented with $\dot{\xi}(0) = -\cos(\theta_0)$, $\dot{\xi}(0) = \sin(\theta_0)$. 

Formulated as an eigenvalue problem, \eqref{ProblemParametricF} can thus be written as:
\begin{subequations}
	\label{ProblemParametricEigenvalue}
	\begin{gather}
		\label{EqsDiff1}
		\dot{\xi} \ddot{\zeta} - \dot{\zeta} \ddot{\xi} = Bo(1 - \zeta)\\
		\label{EqsDiff2}
		\dot{\xi}^2 + \dot{\zeta}^2 = 1
	\end{gather}
	with boundary conditions
	\begin{gather}
		\xi(s = 0) = \zeta(s = 0) = 0 \\
		\dot{\xi}(0) = -\cos(\theta_0), \quad \dot{\zeta}(0) = \sin(\theta_0) \\
		\label{FarUpstream}
		\xi \rightarrow -\infty, \quad \zeta \rightarrow 1 \qquad \text{as}  \qquad s \rightarrow \infty
	\end{gather}
\end{subequations}
where $Bo(\theta_0)$ is the eigenvalue which must be obtained as part of the solution for each value of the contact angle $\theta_0$ of the pinned interface at the wedge's vertex.

\subsection{Solution.}

The problem just formulated can be solved analytically in closed form as shown in Appendix \ref{FisrtApproximationSolution}. Here we will just use a first integral that can be obtained as follows. Combining equations \eqref{EqsDiff1} and \eqref{EqsDiff2} to eliminate $\xi$ gives:
\begin{gather}
	\ddot{\zeta} = -Bo \,(1-\zeta) \sqrt{1-\dot{\zeta}^2},
\end{gather}
which is an ordinary differential equation for the height $\zeta$ of the interface in terms of the arc-length $s$. Integrating it once and matching the solution with the constant thickness $\zeta \rightarrow 1$, $\dot{\zeta} \rightarrow 0$ far upstream from the edge gives the first integral:
\begin{gather}
	\label{FirstIntegral}
	\pm\sqrt{1 - \dot{\zeta}^2} = 1 - \frac{Bo}{2}(1 - \zeta)^2.
\end{gather}

Evaluating it at the triple point, namely $\zeta(s = 0) = 0$, $\pm\sqrt{1 - \dot{\zeta}^2(s=0)} = \cos\theta_0$, gives the Bond number in terms of $\theta_0$, i.e. equation \eqref{BoTheta}:
\begin{gather}
	Bo = \left(\frac{h_\infty}{\ell_c}\right)^2 = 2(1 - \cos\theta_0)
\end{gather}
\begin{figure}[h!]
	\centering
	\includegraphics[width=\linewidth]{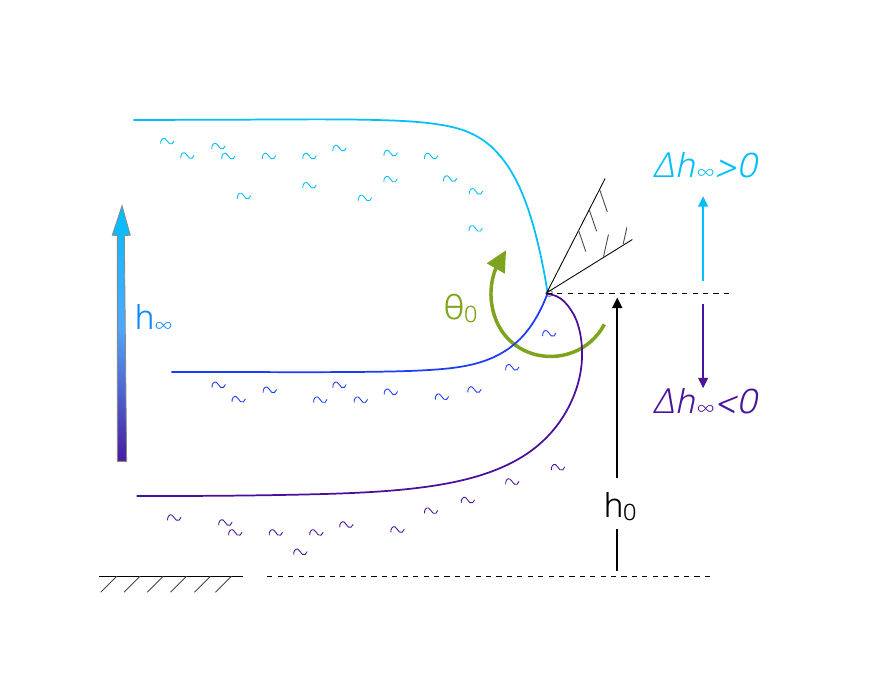}	
	\caption{(Color online) Interface shapes for different values of the effective thickness $\Delta h_\infty = h_\infty - h_0$ of the liquid layer. The thick vertical blue arrow and the thin curved green one indicate the direction of increasing values of $\Delta h_\infty$ and $\theta_0$ respectively. $\theta_0$ range from negative (the two lowest darker blue interfaces) to positive (the highest light blue one) values, with $\theta_0 = 0$ corresponding to a straight horizontal interface at the wedge's vertex height. The arrows also indicate the evolution as the liquid layer is forced to advance against the vertex (green curve starting at $b$ in the $\theta_0$-Bo plane of Figure \ref{BifurcationComplete}), whereas for a layer receding (purple curve starting at $a$ in Figure \ref{BifurcationComplete}) back from the vertex the evolution is reversed.}
	\label{InterfacePlanarTransition}
\end{figure}

In order to keep the reasoning simple we defer the detail analysis of the stability of these equilibrium solutions to Section \ref{AppStability}. However, it can be anticipated that, as common experience shows, the equilibrium configurations are stable for small $\theta_0$ contact angles, although increasing it ultimately results in the destabilization and depinning at the critical $\theta_0 = \pi$, so the solutions corresponding to $\theta_0 > \pi$ are unstable.

We note here that the key point is formulating the problem of the equilibrium of the liquid layer as an the eigenvalue problem \eqref{ProblemParametricEigenvalue}, which naturally leads to the interpretation of Figure \ref{toBo} as the bifurcation diagram of the equilibrium solutions with the eigenvalue $Bo$, or equivalently the liquid layer thickness $h_\infty$, as bifurcation parameter. For each value of the Bond number $Bo < 4$ there are two solutions, the stable branch with $\theta_0 < \pi$, and the unstable one $\theta_0 > \pi $. These two branches merge at $Bo = 4$ and disappear for $Bo > 4$. Therefore, the triple line depinning occurs as a subcritical saddle-node or fold bifurcation.

\section{Receding triple lines.}

The solution of the previous section can be generalized to cases in which the edge height $h_0$ relative to the substrate far upstream is non-zero as shown in Figure  \ref{InterfacePlanarTransition}. Now, in addition to the liquid layer thickness $h_\infty$, the problem has a new length scale $h_0$. However, as shown in Appendix \ref{AppDiffHeights}, it is easy to see that the problem only depends on the difference. Thus, scaling lengths with the effective thickness $\Delta h_\infty = h_\infty - h_0 > 0$ (the negative case is discussed below), leads, to a problem with the exact same form as \eqref{ProblemParametricEigenvalue}, except that now the eigenvalue is the effective Bond number: 
\begin{gather}
	Bo_{eff} = \left(\frac{\Delta h_\infty}{\ell_c}\right)^2 = 2(1 - \cos\theta_0)
\end{gather}

\begin{figure}[h!]
	\includegraphics[width=\linewidth]{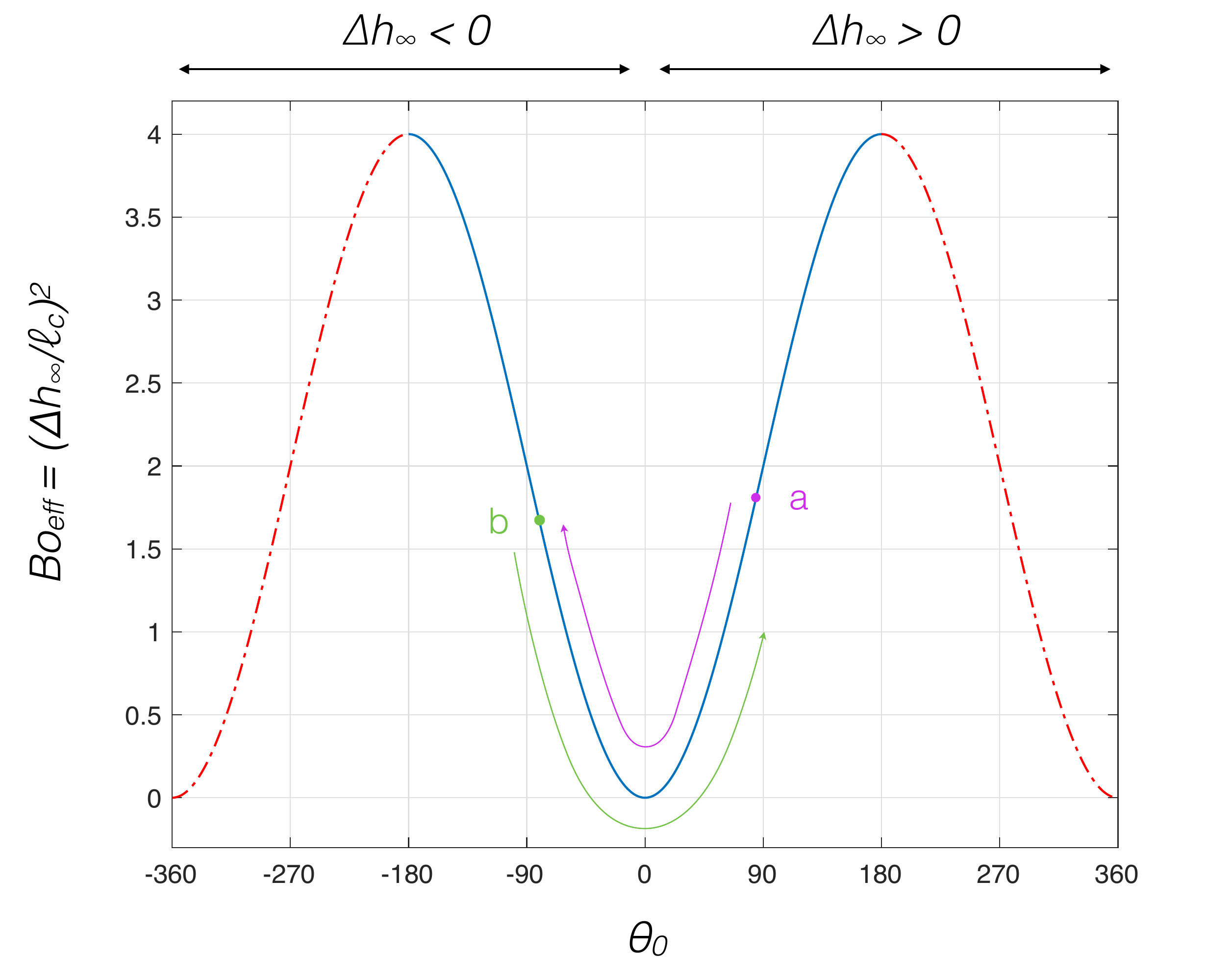}
	\caption{(Color online) The complete bifurcation diagram including positive as well as negative apparent contact angles. The purple arrow, starting at $a$, represents the evolution path for a receding interface, eventually depining at $\theta_0 = -180^\circ$. The green one, starting at $b$, corresponds to the evolution of an advancing liquid front with an initial negative effective thickness $\Delta h_\infty$, eventually depining at $\theta_0 = 180^\circ$. The solution branches in dash-dot lines are inaccessible because they are unstable and deppining occurs before, at $\theta_0 = \pm180^\circ$.}
	\label{BifurcationComplete}
\end{figure}

As suggested in Figure \ref{InterfacePlanarTransition}, this configuration admits both positive and negative values of the effective height $\Delta h_\infty$, when the liquid layer is, respectively, above or below the wedge's vertex. Due to the symmetries of the problem the solutions in this latter case of negative values of $\Delta h_\infty$ (i.e. $h_\infty < h_0$) can be obtained from symmetry considerations by simply flipping the interface about the $\xi$ axis, which changes the sign of $\theta_0$, but leaves the effective Bond number $Bo_{eff}$ unchanged. Therefore, the bifurcation diagram in this generalized case can be obtained from that in Figure \ref{toBo}, extending it to negative values of $\theta_0$ by a reflection about the axis $\theta_0 = 0$ as shown in Figure \ref{BifurcationComplete}.

\subsection{Depinning conditions.}

In order to derive the deppining criterion as the liquid layer recedes, one can consider as initial condition a point in the bifurcation diagram with positive values of $\Delta h_\infty$ and $\theta_0$, as the point $a$ in Figure \ref{BifurcationComplete}. The thickness of the liquid layer, and accordingly the apparent contact angle, can be forced to decrease by removing liquid from far away upstream from the vertex. $\Delta h_\infty$ traverses the equilibrium line $Bo_{eff}(\theta_0)$ in the direction represented by the blue path, reaching eventually $\Delta h_\infty = 0$ for $\theta_0 = 0$, and then becoming negative, forcing accordingly negative apparent contact angles $\theta_0$. The evolution in the physical space is that schematically represented in Figure \ref{InterfacePlanarTransition}, from the light top to the darker blue interfaces. This thinning of the liquid layer can be continued until $\theta_0 = -180^\circ$. Beyond this point it is not possible keep on decreasing $h_\infty$, because $Bo_{eff} = ((h_0 - h_\infty)/\ell_c)^2$ has a local maximum, and therefore $h_\infty$ a local minimum. Consequently, the triple line depins if forced to recede further. The critical deppining condition is thus $Bo_{eff} = 4$, or equivalently $h_\infty = h_0 - 2\ell_c$, and $\theta_0^r = -180^\circ$.

The green path on the other hand, starting at $b$ represents the opposite evolution, namely starting with a liquid layer below the vertex, and therefore pinned with an angle $\theta_0$ below the horizontal. Adding liquid so to increase its thickness will force the growth of $\theta_0$, all the way up through $\theta_0 = 0$, until the limit $\theta_0^a = 180^\circ$, beyond which the triple line depins. The reason is again that $\theta_0^a = 180^\circ$ is a local maximum of $Bo_{eff} = ((h_\infty - h_0)/\ell_c)^2$, and therefore of $h_\infty$, so it is not possible to increase it any further, as the advancing condition being forced would require.

\section{Stability} \label{AppStability}

The heuristic arguments given above showing the loss of stability beyond the critical angles $\theta_0 = \pm180^\circ{}$ can be made rigorous through an analysis of the stability of the equilibrium solutions. We exploit for that purpose the variational structure of the problem, derived for reference in Appendix \ref{VariationalFormulation}. From this point of view, stable solutions correspond to local minima of the energy functional \cite{Guelfand}, i.e. to solutions with the second variation of the energy functional is positive definite. We will proceed in two steps: we first show that the solutions with $|\theta_0| < 180^\circ{}$ are stable; then, we use Turning Point methods\cite{Maddocks, Steen} to prove the loss of stability at $|\theta_0| = 180^\circ{}$, and therefore the instability of the solutions branch corresponding to $\theta_0 > 180^\circ{}$.

\subsection{Stability for $|\theta_0| < 180^\circ{}$.}

We consider here the stability of the equilibrium solutions subject to planar perturbations leaving the interface pinned at the vertex. In addition, we restrict this analysis to apparent contact angles $\theta_0$ smaller than $\theta_0 < 180^\circ{}$. In this case, using the same system of coordinates as in Figure \ref{OverhangingDrop}, the interface $\mathcal{S}$ can be described explicitly in the form $\xi = \xi_{\mathcal{S}}(\zeta)$, with $\xi_{\mathcal{S}}(\zeta=0) = 0$ to enforce pinning at the vertex. This representation is admissible for contact angles in the range $[-180^\circ, 180^\circ]$. Beyond these limits the function $\xi_{\mathcal{S}}(\zeta)$ would be doubled-valued and some more general representation should be used. 

The energy functional becomes thus:
\begin{gather*}
	\mathcal{E} =
	\int_{0}^1{\left\{\sqrt{1 + \xi_{\mathcal{S}}^{'2}(\zeta)} - Bo(1 - \zeta)(\xi_{\mathcal{S}}(\zeta) - H)\right\} d\zeta}
\end{gather*}
where $H$ is a constant. Strictly, $H$ should be let go to infinity, which would make the integral infinite. However, it is easy to check that the value of this constant is immaterial and can actually be set to zero. 
	
We write the perturbed interface as $\xi_{\mathcal{S}}(\zeta) + h(\zeta)$, with the perturbation $h$ bounded and compatible with the pinned interface, that is, $h(0) = 0$, and with a horizontal interface far upstream from the edge, i.e. $h'(\zeta \rightarrow 1) = 0$. Otherwise $h(\zeta)$ is unrestricted, so the perturbed solutions can accommodate variations in the contact angle so $h'(0)$ is arbitrary, although the interface $\xi_{\mathcal{S}}(\zeta) + h(\zeta)$ must still be a well-defined, single valued function of $z$.
	
With this parametrization the second variation of $\mathcal{E}$ is:
\begin{gather*}
	2\delta^2\mathcal{E} = 
	\int_{0}^1{\frac{h'^2}{(1 + \xi_{\mathcal{S}}^{'2})^{3/2}}\, d\zeta}
\end{gather*}
which is clearly non-negative for any admissible perturbation $h(x)$. Thus, the equilibrium solutions found for contact angles $\theta_0$ within $]-180^\circ, 180^\circ[$ are stable under planar pinned perturbations. 

\subsection{Stability for $|\theta_0| \ge 180^\circ{}$.}

The previous analysis shows the stability of the solutions with a contact angle smaller than 180$^\circ{}$. The question of the stability for larger contact angles can be addressed using the elegant results due to Maddocks\cite{Maddocks} for problems with a variational structure as in the present case. In order to in this analysis apparent contact angles larger than $180^\circ{}$ we will use the parametric representation of the interface introduced in the text in Section \ref{Formulation}. The energy functional $\mathcal{E}(u)$ with $u = (\xi, \zeta, \dot{\xi}, \dot{\zeta})$ can be written in that case as:
\begin{subequations}
	\label{VariationalStability}
	\begin{gather}
		\mathcal{E}  = \mathcal{A} + Bo\mathcal{P}
		\label{MinimumPrinciple}
	\end{gather}
	with
	\begin{gather}
		\mathcal{A} = \int_0^\infty{\left(\sqrt{\dot{\xi}^2 + \dot{\zeta}^2}  -1\right)ds} \\
		\mathcal{P} = \int_0^\infty{\left((\zeta - \zeta^2/2)\dot{\xi} + 1/2\right)  ds}
	\end{gather}
	with boundary conditions
	\begin{gather}
		\xi(s = 0) = \zeta(s = 0) = 0 \\
		\dot{\xi}(0) = -\cos(\theta_0), \quad \dot{\zeta}(0) = \sin(\theta_0) \\
		\label{FarUpstream}
		\xi \rightarrow -\infty, \quad \zeta \rightarrow 1 \qquad \text{as}  \qquad s \rightarrow \infty
	\end{gather}
\end{subequations}

\begin{figure}[h!]
	\includegraphics[width=\linewidth]{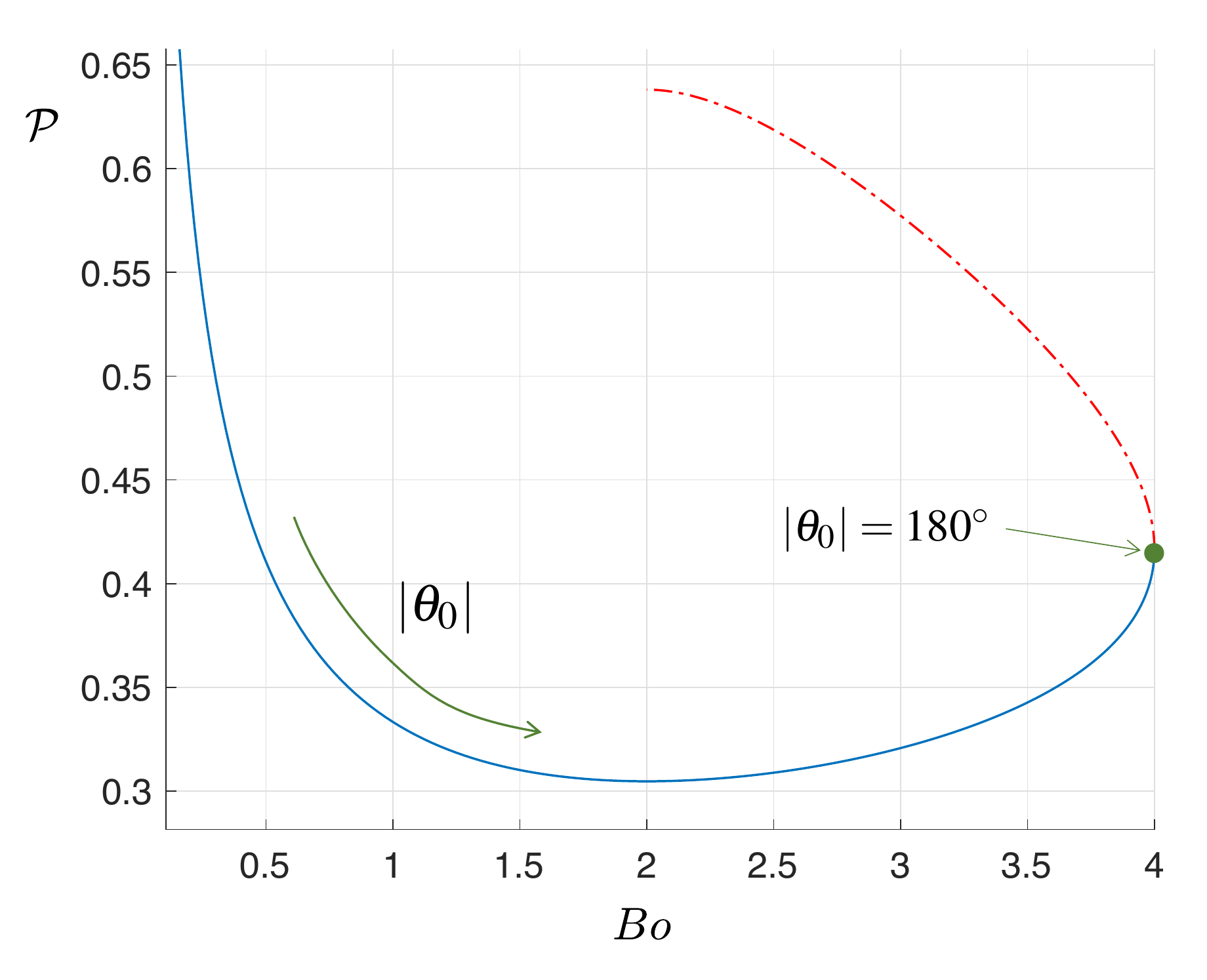}
	\caption{The solid blue and dash-dots red lines represent respectively the stable and unstable equilibrium solutions in the plane $Bo-\mathcal{P}$, with the contact angle as parameter along them. This is the preferred plane\protect\cite{Maddocks} to analyze the stability of the solutions of the problem defined by \eqref{VariationalStability}. The fold ($dBo/d\theta_0 = 0$) at $|\theta_0| = 180^\circ{}$ ($Bo = 4$) represents the loss of stability of the blue stable branch which correspond to $|\theta_0| < 180^\circ{}$.}
	\label{BifurcationMaddocks}
\end{figure}

The Bond number $Bo(\theta_0)$ is an eigenvalue of the problem and the bifurcation parameter. Following Maddocks\cite{Maddocks} and translating his results into our nomenclature, the changes of stability can be studied in the plane $Bo-\mathcal{E}_{Bo}$, where $\mathcal{E}_{Bo} = \mathcal{P}$ is the derivative of the energy functional with respect to the bifurcation parameter $Bo$. In this plane, the equilibrium solutions are represented by curves parametrized in our case by the contact angle $\theta_0$, i.e. $(Bo(\theta_0), \mathcal{P}(\theta_0))$ as shown in Figure \ref{BifurcationMaddocks}. The changes of stability are associated with folds of the solution branches\cite{Maddocks}, defined as points where the derivative $dBo/d\theta_0$ of the bifurcation parameter $Bo$ with respect to $\theta_0$ vanishes, such as the point $\theta_0 = 180^\circ{}$ shown in Figures \ref{BifurcationComplete} and \ref{BifurcationMaddocks}.

This can be seen as follows (see Maddocks\cite{Maddocks} for rigorous proofs). The stable equilibrium solutions correspond to minima of \eqref{MinimumPrinciple}, i.e. to zeros of the first Fr\'echet $u$-derivative $\delta\mathcal{E}$ where the second Fr\'echet $u$-derivative $\delta^2\mathcal{E}$ is non-negative. Since $\delta^2\mathcal{E}$ is a real symmetric quadratic form in the perturbation $(\delta\xi, \delta\zeta, \delta\dot{\xi}, \delta\dot{\zeta})$ of the equilibrium solution, its spectrum is real and form an increasing discrete sequence of eigenvalues. Thus, the second variation being non-negative is equivalent to its critical (smallest) eigenvalue $\mu$ being non-negative as well, and the question of the stability of the equilibrium solutions is reduced to the analysis of the evolution of the critical eigenvalue as the contact angle $\theta_0$ is varied. As shown by Maddocks\cite{Maddocks}, changes of stability can only occur at folds, where the critical eigenvalue crosses zero. This can be easily seen by taking the derivative of the equilibrium condition $\delta\mathcal{E} = 0 $ along the solution branch (i.e. taking the derivative with respect to the contact angle $\theta_0$, denoted with a prime hereafter $d\phi/d\theta_0 = \phi'$). Performing the derivative leads to $\delta^2\mathcal{E} \cdot u' + \delta{\mathcal{E}}_{Bo} \cdot Bo^{'} = 0 $, with $\delta{\mathcal{E}}_{Bo}$ the derivative of $\delta{\mathcal{E}}$ with respect to $Bo$. 
Thus $\delta^2\mathcal{E}$ has a zero eigenvalue when $Bo^{'} = 0 $. This shows that in our problem, changes of stability can only occur at the folds $|\theta_0| = 180^\circ{}$, where $Bo^{'} = 0$ as Figures \ref{BifurcationComplete} and \ref{BifurcationMaddocks} reveal. Furthermore, it can be shown\cite{Maddocks} that the derivative $\mu'$ of the critical eigenvalue with respect to the parameter, $\theta_0$ in this problem, is given, up to a positive factor, by:
\begin{gather}
	\mu' \approx Bo'' \cdot \mathcal{P}'
\end{gather}
where the primes represent the derivatives with respect to the contact angle $\theta_0$. Figure \ref{BifurcationMaddocks} shows that $\mathcal{P}$ is an increasing function of $\theta_0$ around the fold $Bo = 4$, whereas $Bo(\theta_0)$ features a maximum (see also Figure \ref{toBo}), and therefore $Bo''$ is negative. Thus $\mu'$ is negative at the fold. Consequently, since as seen the branch $|\theta_0| < 180^\circ{}$ is stable and therefore the critical eigenvalue $\mu$ is positive, $\mu$ crosses zero at $Bo = 4$ ($\theta =180^\circ{}$) and becomes negative along the branches $|\theta_0| > 180^\circ{}$. This shows therefore that the equilibrium solutions are unstable for apparent contact angles $|\theta_0| > 180^\circ{}$.


\section{Conclusions.}

We have introduced a general methodology capable of providing precise mathematical criteria for depining transitions. In the specific problem analyzed here, and in other cases with similar symmetries, the transition occurs as a subcritical bifurcation, with no equilibrium solutions beyond a critical value of the bifurcation parameter, typically an effective Bond number. This methodology leads to straightforward criteria, which in the analyzed configuration is easily obtained due to the relatively uncomplicated system of ordinary differential equations governing the analyzed 2D configuration. Axisymmetric configurations are still governed by a system of differential equations and precise mathematical depinning conditions, which again arise as subcritical saddle-node bifurcations, can be derived. However, general cases of 3D non-symmetrical configurations lead to free boundary problems involving intricate non-linear partial differential equations, with triple lines that are not known in advanced, but must instead be found as part of the solution. One immediate open question for instance is about the robustness of the saddle-node bifurcation in these more complex problems, i.e. if the saddle-node bifurcation is the generic route to triple line depinning. However, despite these technical difficulties, this new approach seems to uncover an exciting new line of research, which could be helpful towards understanding open problems related to wetting-dewetting transitions, and more broadly the dynamics of free interfaces gliding over solid surfaces. 

\section{Acknowledgments}

We gratefully acknowledge the funding support from NASA award number NNX13AB12A via NASA Kentucky EPSCoR.

\appendix 

\section{Equilibrium solutions.} \label{FisrtApproximationSolution}

As shown in the text, the problem \eqref{ProblemParametricEigenvalue} has the first integral:
\begin{gather}
	\label{FirstIntegralA}
	\pm\sqrt{1 - \dot{\zeta}^2} = 1 - \frac{Bo}{2}(1 - \zeta)^2.
\end{gather}
from which, the problem for $\zeta$ turns to be:
\begin{subequations}
	\label{FinalFirstIntegralA}
	\begin{gather}
		\dot{\zeta} = \pm\sqrt{\frac{Bo}{2}}(1 - \zeta)
			\left( 2 - \frac{Bo}{2}(1 - \zeta)^2\right)^{1/2} \\
		\zeta(s = 0) = 0
	\end{gather}
	where the positive or negative signs apply for contact angles $0 \le \theta_0 \le \pi$ or $\pi \le \theta_0 \le 2\pi$ respectively
\end{subequations}

This equation can be simplified using the auxiliary independent variable $u^2 = 1 - (Bo/4)(1 - \zeta)^2$, which, in the case $0 \le \theta_0 \le \pi$, transforms the problem for $\zeta$ in:
\begin{subequations}
    \begin{gather}
    	\dot{u} = \sqrt{Bo} \, (1-u^2) \qquad (0 \le \theta_0 \le \pi)
    \end{gather}
    to be integrated with
    \begin{gather}
    	u(s = 0) = \left(1-\frac{Bo}{4}\right)^{1/2}
    \end{gather}
\end{subequations}

The integration is straightforward in terms of the hyperbolic trigonometric functions as for instance:
\begin{gather}
	\zeta = 1 - \frac{2}{\sqrt{Bo} \, \cosh\left(\sqrt{Bo} \, s + \tanh^{-1}\sqrt{1 - Bo/4}\right)} \\
	\notag \text {for} \qquad (0 \le \theta_0 \le \pi) 
\end{gather}

In the case of contact angles in the interval $\pi \le \theta_0 \le 2\pi$, the negative sign in equation \eqref{FinalFirstIntegralA} must be chosen initially. However, $\dot{\zeta}$ eventually vanishes at a turning point $s = s_{tp} = Bo^{-1/2}\tanh^{-1}\sqrt{1 - Bo/4}$, corresponding to the minimum negative height of the drop. The integration then proceeds for $s > s_{tp}$ with positive values of $\dot{\zeta}$, so the sign in \eqref{FinalFirstIntegralA} must be switched back to plus. The solution can be written thus as:
\begin{subequations}	
	\begin{gather}
		\zeta = 1 - \frac{2}{\sqrt{Bo} \, \cosh\left(-\sqrt{Bo} \, s + \tanh^{-1}\sqrt{1 - Bo/4}\right)}  \\
	\notag \text {for} \qquad  (\pi \le \theta_0 \le 2\pi) 
	\end{gather}
	
	On the other hand, the equation for $\xi$ follows from \eqref{FirstIntegralA} as:
	\begin{gather}
		\dot{\xi} = -1 + \frac{Bo}{2}(1 - \zeta)^2.
	\end{gather}
	to be integrated with the condition $\xi(s = 0) = 0$. Thus, using the previous solution for $\zeta$ gives:
	\begin{gather}
		\xi = -s + \frac{2}{\sqrt{Bo}}\left[\tanh\left(\sqrt{Bo}(s+s_{tp}\right) - \sqrt{\left(1 - \frac{Bo}{4}\right)}\right] \\
	\notag \text {for} \qquad  (0 \le \theta_0 \le \pi) \\
		\xi = -s - \frac{2}{\sqrt{Bo}}\left[\tanh\left(\sqrt{Bo}(-s+s_{tp}\right) - \sqrt{\left(1 - \frac{Bo}{4}\right)}\right] \\
	\notag \text {for} \qquad  (\pi \le \theta_0 \le 2\pi)
	\end{gather}
\end{subequations}

\section{Variational formulation.}\label{VariationalFormulation}

The system \eqref{ProblemParametricEigenvalue} used in the text can be as well obtained as the Euler-Lagrange equations\cite{Guelfand} for the shape that minimizes the total energy \cite{LandauFluids, Steen} $E = \int{\{\sigma dA +  PdV\}}$, including the surface and the pressure-gravitational potential energies. In the specific case of the 2D configuration being studied, it is more convenient to consider the energy per unit length (normal to the paper in Figure \ref{OverhangingDrop}), so the volume integral of the potential energy becomes a surface integral extended to the area enclosed between the interface and the substrate, whereas the surface integral reduces to a line integral along the interface. Assuming a uniform surface tension along the interface, and writing $\mathcal{E}$ for the energy per unit length scaled with $\sigma h_\infty$, the appropriate energy functional can be written as the line integral:
\begin{gather}
	\label{EnergyFunctionalApp}
	\mathcal{E} = 
	\int_0^\infty{\left(\sqrt{\dot{\xi}^2 + \dot{\zeta}^2} + Bo(\zeta - \zeta^2/2)\dot{\xi} - \mathcal{E}_0\right) ds}
\end{gather}
where $ \mathcal{E}_0 = 1 - Bo/2$ is a constant added to render the integral convergent as $s \rightarrow \infty$, where $\dot{\xi} \rightarrow -1$, $\zeta \rightarrow 1$. However this constant plays no role in the solution and can be dropped.

Standard methods give the system:
\begin{subequations}
	\label{VariationalFormApp}
	\begin{gather}
		\ddot{\xi} = -Bo \, \dot{\zeta} \, (1-\zeta) \\
		\ddot{\zeta} = Bo \, \dot{\xi} \, (1-\zeta)
	\end{gather}
\end{subequations}
which can also easily be derived from Eq. \ref{ProblemParametricF}.

It can be noticed that integral of the term $\zeta\dot{\xi}$ in \eqref{EnergyFunctionalApp} represents the volume (per unit length) of the liquid layer. $Bo$ is therefore the Lagrange multiplier of the constrained problem of finding the shape of minimum surface area (per unit length) subject to a constant volume (per unit area). In general cases, the Lagrange multiplier is not known in advance and must be obtained as part of the solution, so in these cases it is more advantageous to formulate the problem as that of finding the interface shape of minimum surface area constrained to a constant volume. However, due to the special configuration of our problem, namely with a constant liquid thickness far upstream, the multiplier can be determined beforehand and the problem can thus be formulated as the unconstrained minimization of the energy functional \eqref{EnergyFunctionalApp}.

\section{Triple line pinned at a non-zero height.} \label{AppDiffHeights}

We derive here the formulation of the problem to find the equilibrium shape of the interface in cases where the height of the edge where the interface is pinned is not the same as that of the horizontal substrate far upstream. We use the formulation introduced in the previous Appendix, based on a variational approach, which is equivalent to \eqref{ProblemParametricEigenvalue} used in the text. The origin of coordinates is again chosen at the wedge's vertex where the interface is pinned. The effective height of the liquid layer far upstream from the triple line is therefore $\Delta h_\infty = h_\infty - h_0 > 0$ (we consider below the case of negative values of $\Delta h_\infty$), where the absolute heights, $h_\infty$ and $h_0$, are measured relative to some horizontal reference, for instance the substrate far upstream. The pressure distribution is thus given by:
\begin{gather}
	p - p_\infty = \rho g (\Delta h_\infty - z),
\end{gather}
Proceeding as before, lengths are made dimensionless with the effective liquid layer thickness $\Delta h_\infty$, whereas the Bond number is defined now in terms of $\Delta h_\infty$, namely $Bo_{eff} = \rho g \Delta h_\infty^2/\sigma$. Using the more symmetrical form of the differential equations derived from the variational formulation \eqref{VariationalFormApp}, the shape of the interface is thus given by the solution of an eigenvalue problem with the form, exactly equivalent to that in the main text	 \eqref{ProblemParametricEigenvalue}:
\begin{subequations}
	\label{ProblemPositiveDiff}
	\begin{gather}
		\label{ProblemParametricEigenvalueDiffEqXi}
		\ddot{\xi} = -Bo_{eff} \, \dot{\zeta} \, (1 - \zeta) \\
		\label{ProblemParametricEigenvalueDiffEqZeta}
		\ddot{\zeta} = Bo_{eff} \, \dot{\xi} \, (1 - \zeta)
	\end{gather}
	with boundary conditions
	\begin{gather}
		\xi(s = 0) = \zeta(s = 0) = 0 \\
		\dot{\xi}(0) = -\cos(\theta_0), \quad \dot{\zeta}(0) = \sin(\theta_0) \\
		\label{FarUpstream}
		\xi \rightarrow -\infty, \quad \zeta \rightarrow 1 \qquad \text{as}  \qquad s \rightarrow \infty
	\end{gather}
\end{subequations}

The solution of this problem is therefore the same as that for zero height the wedge's edge, but using now as eigenvalue the effective Bond number, given by:
\begin{gather}
	\left(\frac{h_\infty - h_o}{\ell_c}\right)^2 = Bo_{eff} = 2(1 - \cos\theta_0)
\end{gather}
In particular, the bifurcation diagram remains unchanged.

\subsection{Negative heights} 
This configuration admits as well negative values of the relative height $\Delta h_\infty$ as shown in Figure \ref{InterfacePlanarTransition}. The solution in these cases can be obtained proceeding similarly. Using as length scale the absolute value $|\Delta h_\infty|$ of the relative height leads to
\begin{subequations}
	\begin{gather}
		\label{ProblemParametricEigenvalueDiffEqXi}
		\ddot{\xi} = -Bo_{eff} \, \dot{\zeta} \, (-1 - \zeta) \\
		\label{ProblemParametricEigenvalueDiffEqZeta}
		\ddot{\zeta} = Bo_{eff} \, \dot{\xi} \, (-1 - \zeta)
	\end{gather}
	subject to
	\begin{gather}
		\xi(s = 0) = \zeta(s = 0) = 0 \\
		\dot{\xi}(0) = -\cos(\theta_0), \quad \dot{\zeta}(0) = \sin(\theta_0) \\
		\label{FarUpstream}
		\xi \rightarrow -\infty, \quad \zeta \rightarrow -1 \qquad \text{as}  \qquad s \rightarrow 
\infty
	\end{gather}
\end{subequations}

This problem can be brought back to the same form as (\ref{ProblemPositiveDiff}) with the change of variables $\zeta \rightarrow -\zeta$, $\theta_0 \rightarrow -\theta_0$, which represent the reflection about the $\xi$ axis of the solutions $[\xi(s), \zeta(s)]$ corresponding to positive values of $\Delta h_\infty$, i.e. a reflection about the horizontal line through the vertex of the wedge as anticipated. $Bo_{eff}$ has been left unchanged, and therefore it must be an even function of $\theta_0$, which means the bifurcation diagram is even in $\theta_0$ as represented in Figure \ref{BifurcationComplete}.

This derivation could have been based on the appropriate energy functional in the case of negative heights, namely
	\begin{gather}
		\label{EnergyFunctionalNegative}
		\mathcal{E} = 
		\int_0^\infty{\left(\sqrt{\dot{\xi}^2 + \dot{\zeta}^2} \right.
		\left. - Bo(\zeta - \zeta^2/2)\dot{\xi} - 1 + (3/2)Bo\right) ds}
	\end{gather}
which accounts for the opposite sign (convexity) of the interface curvature in these cases of negative height.

%

\end{document}